%
\let\useblackboard=\iftrue
%
%
\newfam\black
\input harvmac.tex
%
\input epsf.tex
\ifx\epsfbox\UnDeFiNeD\message{(NO epsf.tex, FIGURES WILL BE
IGNORED)}
\def\figin#1{\vskip2in}
\else\message{(FIGURES WILL BE INCLUDED)}\def\figin#1{#1}\fi
\def\ifig#1#2#3{\xdef#1{fig.~\the\figno}
\midinsert{\centerline{\figin{#3}}%
\smallskip\centerline{\vbox{\baselineskip12pt
\advance\hsize by -1truein\noindent{\bf Fig.~\the\figno:} #2}}
\bigskip}\endinsert\global\advance\figno by1}
\noblackbox
\def\Title#1#2{\rightline{#1}
\ifx\answ\bigans\nopagenumbers\pageno0\vskip1in%
\baselineskip 15pt plus 1pt minus 1pt
\else
\def\listrefs{\footatend\vskip
1in\immediate\closeout\rfile\writestoppt
\baselineskip=14pt\centerline{{\bf
References}}\bigskip{\frenchspacing%
\parindent=20pt\escapechar=` \input
refs.tmp\vfill\eject}\nonfrenchspacing}
\pageno1\vskip.8in\fi \centerline{\titlefont #2}\vskip .5in}
 
scaled\magstep3
 
scaled\magstep3
 
scaled\magstep3
 
scaled\magstep3
 
scaled\magstep3
\ifx\answ\bigans\def\tcbreak#1{}\else\def\tcbreak#1{\cr&{#1}}\fi

\useblackboard
\message{If you do not have msbm (blackboard bold) fonts,}
\message{change the option at the top of the tex file.}

\font\blackboard=msbm10 scaled \magstep1
\font\blackboards=msbm7
\font\blackboardss=msbm5
\textfont\black=\blackboard
\scriptfont\black=\blackboards
\scriptscriptfont\black=\blackboardss

\else

\fi
%

%
\def\yboxit#1#2{\vbox{\hrule height #1 \hbox{\vrule width #1
\vbox{#2}\vrule width #1 }\hrule height #1 }}
\def\fillbox#1{\hbox to #1{\vbox to #1{\vfil}\hfil}}
\def\ybox{{\lower 1.3pt \yboxit{0.4pt}{\fillbox{8pt}}\hskip-0.2pt}}

\def\comments#1{}

\def\vev#1{\langle{#1}\rangle}

\def\II{\relax{I\kern-.07em I}}

\def\inbar{\,\vrule height1.5ex width.4pt depth0pt}
\def\IZ{\relax\ifmmode\mathchoice
{\hbox{\cmss Z\kern-.4em Z}}{\hbox{\cmss Z\kern-.4em Z}}
{\lower.9pt\hbox{\cmsss Z\kern-.4em Z}}
{\lower1.2pt\hbox{\cmsss Z\kern-.4em Z}}\else{\cmss Z\kern-.4em
Z}\fi}
\def\IB{\relax{\rm I\kern-.18em B}}
\def\IC{{\relax\hbox{$\inbar\kern-.3em{\rm C}$}}}
\def\ID{\relax{\rm I\kern-.18em D}}
\def\IE{\relax{\rm I\kern-.18em E}}
\def\IF{\relax{\rm I\kern-.18em F}}
\def\IG{\relax\hbox{$\inbar\kern-.3em{\rm G}$}}
\def\IGa{\relax\hbox{${\rm I}\kern-.18em\Gamma$}}
\def\IH{\relax{\rm I\kern-.18em H}}
\def\IK{\relax{\rm I\kern-.18em K}}
\def\IP{\relax{\rm I\kern-.18em P}}
\def\pp{{\relax{=\kern-.42em |\kern+.2em}}}

\font\cmss=cmss10 \font\cmsss=cmss10 at 7pt
\def\IR{\relax{\rm I\kern-.18em R}}

\def\frac#1#2{{{#1} \over {#2}}}

%
%

\def\PR{{\it Phys. Rev.\ }}

\def\JMP{{\it J. Math. Phys.\ }}

\Title{ \vbox{\baselineskip12pt\hbox{hep-th/9907062}
\hbox{BROWN-HET-1189}
}}
{\vbox{
\centerline{Comments on a Covariant Entropy Conjecture}}}

\centerline{ David A. Lowe}
\medskip

\centerline{Department of Physics}
\centerline{Brown University}
\centerline{Providence, RI 02912, USA}
\centerline{\tt lowe@het.brown.edu}

\medskip

\centerline{\bf{Abstract}}

\noindent
Recently Bousso conjectured the entropy crossing a certain light-like
hypersurface is bounded by the surface area. We point out a number of
difficulties with this conjecture.

\vfill
\Date{\vbox{\hbox{\sl July, 1999}}}

\lref\bousso{R. Bousso, ``A Covariant Entropy Conjecture,''
hep-th/9905177.}
\lref\page{D.N. Page, ``Comment on a universal upper bound on the
entropy-to-energy ratio for bounded systems,'' \PR {\bf 26} (1982)
947.}
\lref\beke{J.B. Bekenstein, \PR {\bf D23} (1981) 287.}
\lref\gsl{J.B. Bekenstein, \PR {\bf D7} (1973) 2333; \PR {\bf D9}
(1974) 3292.}
\lref\bekenew{J.B. Bekenstein, ``Non-Archimedian character of quantum
buoyancy and the generalized second law,'' gr-qc/9906058.}
\lref\pelath{M.A. Pelath and R.M. Wald, ``Comment on entropy bounds
and the generalized second law,'' gr-qc/9901032.}
\lref\unruh{W.G. Unruh and R.M. Wald, \PR {\bf D25} (1982) 942;
\PR {\bf D27} (1983) 2271.}
\lref\thooft{G. 't Hooft, ``Dimensional reduction in quantum
gravity,'' gr-qc/9310026.}
\lref\susskind{L. Susskind, ``The world as a hologram,'' \JMP {\bf 36} 
(1995) 6377, hep-th/9409089.}
\lref\easther{R. Easther and D.A. Lowe, ``Holography, cosmology and
the second law of thermodynamics,'' hep-th/9902088.}
\lref\venez{G. Veneziano, ``Pre-bangian origin of our entropy and time 
arrow,'' hep-th/9902126.}
\lref\bak{D. Bak and S.-J. Rey, ``Cosmic holography,''
hep-th/9902173.}
\lref\linde{N. Kaloper and A. Linde, ``Cosmology vs. holography,''
hep-th/9904120.} 
\lref\brustein{R. Brustein, ``The generalized second law of
thermodynamics in cosmology,'' gr-qc/9904061.}
\lref\larus{D.A. Lowe and L. Thorlacius, ``AdS/CFT and the Information
Paradox,'' hep-th/9903237.}
\lref\laru{L. Susskind and L. Thorlacius, ``Gedanken Experiments
Involving Black Holes,'' \PR {\bf D49} (1994) 966, hep-th/9308100.}
\lref\fischler{W. Fischler and L. Susskind, ``Holography and
Cosmology,'' hep-th/9806039.}

\newsec{Covariant Entropy Conjecture}

Recently Bousso \bousso\ made the interesting conjecture that the entropy $S$
passing through a certain hypersurface bounded by a two-dimensional 
spatial surface $B$ with area $A$ must satisfy the bound 
\eqn\bound{
S \leq A/4~.
}
The hypersurface $L$ in question is to be generated by one of the four 
null congruences orthogonal to $B$, with non-positive expansion in
the direction away from $B$. The matter fields in the theory are
required to satisfy the dominant energy condition.
This {\it covariant entropy conjecture} 
is motivated by the proposed holographic principle of
't Hooft and Susskind \refs{\thooft,\susskind}, 
and recent work on attempts to generalize this
principle to cosmological backgrounds \refs{\fischler\easther \venez \bak
\linde{--}\brustein}.

In this note, we point out a number of difficulties with this
proposal. We begin by noting the choice of units is such that
the fundamental constants satisfy $\hbar=c=G=k=1$. 
With appropriate factors of $G$ and $\hbar$ restored the bound 
becomes $S \leq A/(4 \hbar G)$, and we see the bound becomes trivial
in the classical limit, for fixed gravitational interactions. 
If the bound is to hold at all, it must hold in the full
theory of quantum gravity coupled to matter. 

However, as is well-known, the dominant energy condition (and also the 
weak energy condition) fails even
for free quantum fields. One consequence of the dominant energy
condition is that the local energy density appears positive definite.
If one computes the expectation
value of the normal-ordered energy momentum tensor 
$\vev{\psi |T_{\mu\nu} | \psi}$ in free scalar field theory, for a
state $\psi$ that is an admixture of the ground state and a
two-particle state, interference terms in the expectation value can
lead to negative values for the energy density. 
As it stands therefore, the conjecture is
inconsistent. If one insists on taking the classical limit, the bound
becomes trivial. As soon as one goes to the quantum theory, the
conditions for the bound to hold are violated, 
except possibly for a theory with no matter
content. 

One reason for imposing the dominant energy
condition was to rule out the possibility of superluminal entropy 
flow, which would allow for easy violations of the bound, as we see in 
a moment. 
In addition, one wants to rule out creating large amounts of entropy
with little energy, by simultaneously
creating matter with positive and negative
energy density.
Of course
one could try to replace the dominant energy condition with a weaker
constraint. However the following example illustrates difficulties
with the bound that are independent of this condition.

\ifig\fone{Penrose diagram for a black hole. $B$ is a surface on the
event horizon. $L$ is a light-like hypersurface with zero expansion, bounded by $B$.}
{\epsfysize=3in\epsfbox{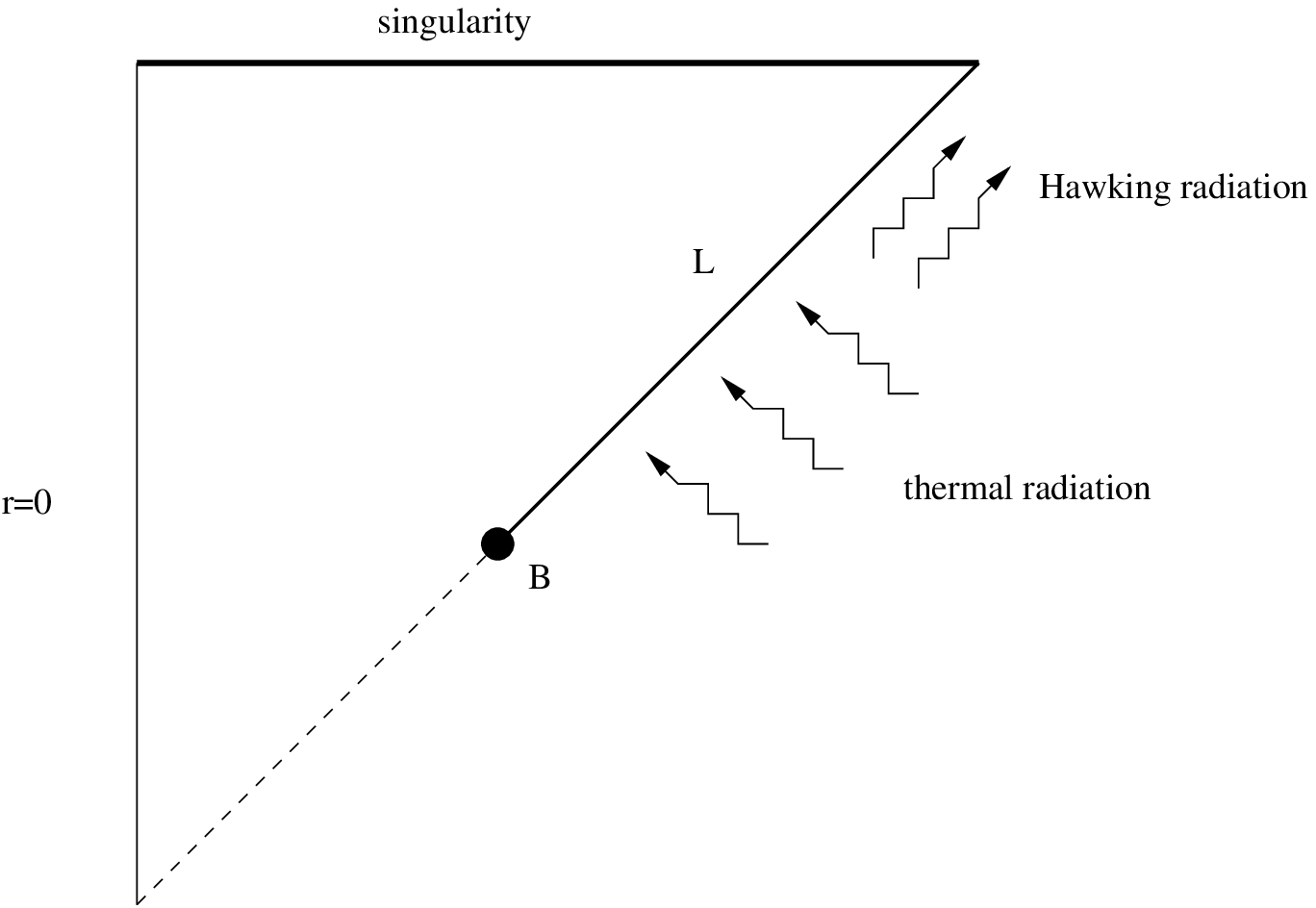}}

Consider a black hole in equilibrium with thermal radiation at the
same temperature as the Hawking temperature. As discussed in \bousso,
one can take $B$ to be the event horizon at some time, and construct
the hypersurface $L$ using future-directed 
outgoing null generators, as shown in
\fone. Since the black hole's evaporation is supported by the
ingoing thermal radiation, the geometry near the horizon is
static. One therefore has an infinite amount of time to send a
constant flux of entropy across the hypersurface $L$, in violation of
the bound. To be precise, we define the entropy crossing $L$ to
be the proper entropy flux, integrated over $L$.
This problem is independent of the matter content of
the theory (and hence any energy conditions one might
choose to impose), since even for pure gravity, a quantum black hole will
Hawking radiate gravitons. Note also that for a large black hole the
geometry is such that caustics need not force the
hypersurface $L$ to approach the singularity.

It is interesting to consider how this example is consistent with the
generalized second law of black hole thermodynamics. While it is true
the black hole absorbs an infinite amount of entropy as time goes to
infinity, it emits an equal (or larger) 
amount of entropy in the form of Hawking
radiation, in accord with the second law. However the entropy emitted
cannot cross the hypersurface $L$ in a causal way. 
The wavelength of Hawking particles is of order the size 
of the black hole. They are best thought of as originating
from outside the black hole, of order the Schwarzschild radius from the
event horizon. The Hawking particles themselves do not contribute to the proper
entropy passing through $L$, since they are undetectable to a freely
falling observer as she crosses the horizon.

If an arbitrarily large amount of entropy can cross the hypersurface
$L$, how can one regard the $\log$ of the 
number of internal states of the black
hole as $A/4$? We will analyze this question in the scenario for the
resolution of the black hole information problem discussed in \larus.
Roughly speaking, in this picture information crosses the
horizon in a completely causal manner, but is effectively transferred
to the Hawking radiation in a non-local way as it hits the singularity. 
Thus we do indeed have super-luminal propagation of entropy in this
picture. This component of the entropy 
does not contribute to the local entropy flux
passing through $L$. 
Although an arbitrarily large
amount of entropy does cross $L$, a low-energy observer inside
the black hole could never detect more entropy than $A/4$. In order
for an observer inside to live long to detect more entropy than $A/4$
she would have to undergo a trans-Planckian acceleration of order
$e^{M^2}$. One sees this via similar gedanken experiments to ones
considered in \refs{\larus,\laru}. Likewise an observer entering the horizon at
late times will see that most of the entropy has already hit the
singularity, preserving the bound on the number of observable internal states.

\ifig\ftwo{Penrose diagram for collapsing spherical dust cloud.
 $L$ is light-like hypersurface that intersects all the dust cloud. In 
order to avoid caustics, $L$ will be deformed to lie along the dotted line.}
{\epsfysize=3in\epsfbox{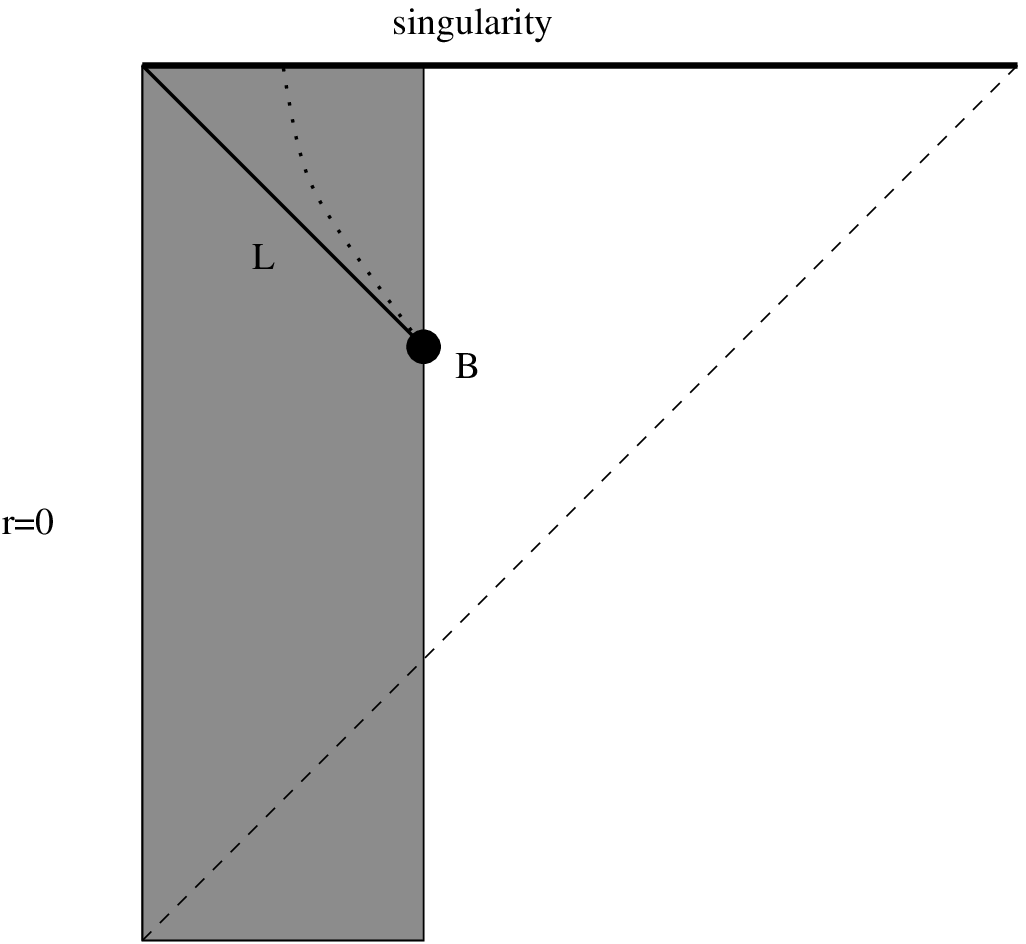}}

Having presented perhaps the clearest counterexample to the bound
\bound, we now consider a number of other objections to the arguments
Bousso presents as evidence for the bound.
Consider a collapsing dust cloud, and construct the hypersurface $L$
as indicated in \ftwo, so that it intersects the whole dust cloud. 
Bousso's claim is that caustics will force
$L$ to take a more circuitous route to the singularity, in such a way
that the surface $L$ does not intersect the whole dust cloud,
preserving the bound \bound. This
relies on the fact that a highly entropic system can never be
spherically symmetric so many caustics will be present. Here we simply 
point out that in the semiclassical approximation, one requires only
that the expectation value of the energy momentum tensor be
spherically symmetric to avoid caustics. This does not provide a
significant constraint on the entropy of such configurations. 
Beyond the semiclassical
approximation, the null convergence condition does not hold, so such
caustics need not form in the first place. Furthermore, since the
construction of the surface $L$ is formulated in terms of classical
geometric quantities, it is not clear how Bousso's construction
carries over to the full quantum theory.

Much of the other evidence Bousso presents for the bound involves
showing consistency with Bekenstein's conjectured bound \beke
\eqn\bekeb{
S\leq 2\pi E R~,
}
in a number of different examples. Here $E$ is the total energy of the 
system, and $R$ is the circumferential radius, defined as
$R=\sqrt{A/4\pi}$ with $A$ the area of the smallest sphere surrounding 
the system.
This bound has been much discussed since its original proposal \beke.
The bound appears to hold for large
systems, provided the number of matter species is small
\refs{\page, \bekenew}.
It can easily be violated for sufficiently small systems (for
example
the free
scalar field case already mentioned), for a large
number of matter species \page, or for systems at sufficiently low
temperature. Likewise,
one can take a large number of copies of a small system, to make a
large system that violates \bekeb.
In general, the generalized second law \gsl\ does not imply the bound
\bekeb\ 
\refs{\unruh, \pelath}.
Some special systems for which \bekeb\ is violated can be used to
construct counterexamples to the covariant entropy bound.

For instance consider a normal region (i.e. not trapped or
anti-trapped) in a
Friedmann-Robertson-Walker cosmology, as discussed in \bousso. 
The covariant entropy bound
implies that the entropy on a spatial hypersurface inside the apparent 
horizon with radius $r_{AH}$, should satisfy $S\leq \pi r_{AH}^2$. This
follows from the bound \bekeb\ if this region is treated as a
Bekenstein system \bousso. Thus the bound  \bound\ can potentially 
be violated for systems
which violate \bekeb. The simplest example of such a system is a gas
of $N$ species of free particles in a box of size $R$ with energy
$E$. The entropy of this gas will diverge as $\log N$. One cannot use
this fact to constrain the value of $N$, as it simply means the bound
\bound\ 
is not a universal bound for any system of matter coupled to gravity.

\newsec{Conclusions}

We have noted a number of difficulties with the current formulation of 
the covariant entropy conjecture.  
We propose that in general backgrounds, the only unexpected
entropy bounds arise from demanding validity of the generalized second 
law, as suggested in \easther. This law has passed a number of highly
nontrivial consistency checks \unruh. However whether the second law gives
rise to holographic style bounds is system dependent. It does not
constrain the entropy density in the early universe, nor in the final
phase of a recollapsing universe. However if an isolated system can collapse 
to a black hole, the second law implies 
the entropy satisfies $S\leq A/4$.

\bigskip
{\bf Acknowledgments}

I thank Richard Easther, Sanjaye Ramgoolam and 
Don Marolf for helpful discussions.
The research of D.L. is supported in part by DOE grant DE-FE0291ER40688-Task A.

\listrefs
\end